*CORRESPONDENCE
Jannik Henze
✉ jannik.henze@uni-koeln.de

# AI-supported data analysis boosts student motivation and reduces stress in physics education

Jannik Henze[1]*, Julia Lademann[2], Sebastian Becker-Genschow[2] and André Bresges[1]

[1]Institute for Physics Education, University of Cologne, Cologne, Germany, [2]Digital Education Research, Faculty of Mathematics and Natural Sciences, University of Cologne, Cologne, Germany

The integration of artificial intelligence (AI) into education presents new opportunities for supporting learning processes. This study investigates the impact of AI-assisted versus traditional *Excel*-based data analysis on both learning outcomes and emotional-motivational responses in a physics education context. A custom *GPT*-based chatbot, *ExperiMentor*, was developed to support student teachers in analyzing experimental data from thread and spring pendulum experiments. Fifty student teachers were randomly assigned to either the AI or *Excel* group, with both groups completing identical tasks in a guided setting. Learning progress was measured using pre- and post-tests, while emotional and motivational variables were assessed through structured surveys. Both groups demonstrated significant learning gains, with no statistically significant differences found between them in terms of cognitive performance. However, the AI group reported substantially higher levels of engagement, enjoyment, and perceived method effectiveness compared to the *Excel* group. These findings suggest that interactive AI tools may enhance the affective dimensions of learning, even when cognitive outcomes remain comparable to traditional methods. The results underscore the importance of integrating AI not as a replacement for instructional design, but as a supportive element within pedagogical frameworks. Future research should explore long-term retention effects, the role of learner diversity, and comparisons with other forms of pedagogical support.

KEYWORDS

**artificial intelligence, data analysis, excel, learning, motivation, physics education**

## 1 Introduction

The role of digital technologies in education is expanding, with artificial intelligence (AI) attracting particular interest as a potential support for complex learning processes. Across disciplines, educators and researchers are exploring how AI tools might assist students in developing not only subject-specific knowledge, but also broader skills for navigating increasingly data-rich and cognitively demanding learning environments (Roll and Wylie, 2016; Winkelmann et al., 2021). As educational strategies evolve toward adaptability, the integration of AI emerges as a pivotal factor at both school and university levels that is redefining the boundaries of teaching and learning (Kuleto et al., 2021; Bacia et al., 2024).

One of the most widely discussed developments in this context is the emergence of Large Language Models (LLMs), such as *ChatGPT*. In the context of this study, the term AI primarily refers to LLMs such as *ChatGPT*, which represent a specific subset of AI focused on natural language processing.

In physics education, LLMs can be used to support tasks like problem solving or interpreting experimental data. Studies suggest these models can reduce barriers to engagement by making learning more flexible and responsive (Küchemann et al., 2024).

Furthermore *ChatGPT* can simplify the process of tackling complex physics problems by offering step-by-step explanations, providing a conversational partner and adapting to different learning styles, making physics more accessible (Tong et al., 2024). The growing capabilities of AI in processing and interpreting data have opened up new possibilities for integrating these tools into physics instruction (Farrokhnia et al., 2024).

Despite the availability of AI tools such as *ChatGPT*, students often struggle to apply them productively, especially when lacking guidance or a solid understanding of underlying physics concepts (Kechel and Wodzinski, 2015; Low and Kalender, 2023). Therefore, explicit instructional strategies are necessary to help students bridge conceptual gaps and improve their ability to leverage AI effectively.

These technological shifts necessitate a critical evaluation of educators' roles and pedagogical methodologies (Popenici and Kerr, 2017). As AI-powered applications gain growing importance in educational strategies (Ständige Wissenschaftliche Kommission der Kultusministerkonferenz, 2024) at school and university levels institutions must develop robust frameworks to effectively incorporate this technology, addressing specific challenges such as technological complexity, data privacy concerns, ethical considerations and other potential barriers to equal learning opportunities.

Given the rapid pace of development and ongoing debate, it is crucial to examine not only what AI tools can offer, but also where their limits lie. While AI-based tools show potential, concerns remain regarding their accuracy, potential bias, and long-term pedagogical value. This study conducts a comprehensive analysis through a comparison of two learning tools used in a physics context. A custom GPT-based chatbot called *ExperiMentor* was developed to support students in analyzing data from physics experiments. In a controlled setting, one group of student teachers, university students studying to become teachers after their master's degree, used *ExperiMentor* to evaluate pendulum experiments, while a second group used *Microsoft Excel*. Both groups received the same tasks, allowing for a comparison of their learning outcomes as well as their emotional and motivational experiences.

The study focuses on two key questions: first, how do students' learning outcomes change within each group, and how do the groups compare? Second, how do students experience each tool emotionally and motivationally, in terms of engagement, frustration, and perceived effectiveness? These questions, derived from a self-developed questionnaire, explore the conditions under which AI tools might support student learning, and where challenges or limitations may arise. The paper is structured as follows: Section 2 reviews current research on AI in education. Section 3 outlines the study design and methods. Section 4 presents the findings, followed by a discussion in Section 5. Section 6 concludes with implications and directions for future research.

# 2 State of research

## 2.1 Educational potential and technical foundations

Recent developments in AI research have drawn attention to the educational potential of LLMs. To contextualize the present study, this section first outlines how LLMs such as *ChatGPT* function and why they are gaining relevance in educational discourse. From a technical perspective, LLMs like *ChatGPT* are trained to predict the next token (e.g., word or character) in a sequence, using massive datasets to model patterns in language (Bender et al., 2021; Ouyang et al., 2022). *ChatGPT*-4, the model used in this study, is multimodal and can process both text and image inputs. It was trained to improve its factual accuracy and responsiveness (OpenAI et al., 2023) compared to predecessor models. The educational potential of such technology has led to growing research interest (Roll and Wylie, 2016; Chen et al., 2020; Mahligawati et al., 2023) along with calls for educators to understand and engage AI in order to prepare for future teaching and learning contexts (Zawacki-Richter et al., 2019). Current findings remain often preliminary, diverse and context dependent. There is no clear consensus yet on the long-term impact or optimal use of AI tools in education. Nevertheless studies show that ChatGPT can increase learning gains (Alarbi et al., 2024; Kestin et al., 2025) and enhance engagement and motivate learners (Wang and Fan, 2025). AI integration in education has mostly taken place at the individual level, not institutionally (Vincent-Lancrin and van der Vlies, 2020). As a result, teacher training is considered crucial for meaningful implementation (Salas-Pilco et al., 2022).

## 2.2 Applications and opportunities in physics education

Beyond their technical foundations, LLMs have increasingly been applied in subject-specific domains. Studies suggest that AI tools like *ChatGPT* support interactive, student-centered learning by offering real-time explanations and adapting to individual input, particularly in explaining complex physics concepts and problem-solving strategies (Popenici and Kerr, 2017; Lin et al., 2022; Liang et al., 2023; Santos, 2023; Küchemann et al., 2024). However, it remains unclear how effective this feedback is compared to traditional forms (Farrokhnia et al., 2024).

Previous studies suggest that *ChatGPT* may be a useful assistant tool, surpassing traditional tools such as *Microsoft Excel* in certain tasks (Halaweh, 2023) while also providing students with the advantage of actively engaging with educational materials, allowing a flexible learning environment (Padma and Rama, 2022; La Ossa et al., 2024; Ng et al., 2024). These capabilities have been studied in relation to their potential to support structured reasoning and dataset interpretation (Mustofa et al., 2024). In one study, *ChatGPT-4* outperformed engineering students on the Force Concept Inventory (FCI), suggesting its potential for physics instruction (Kieser et al., 2023).

## 2.3 Limitations, accuracy and concerns

Despite these applications, recent research also highlights important limitations of LLMs in educational contexts, particularly in terms of accuracy, reliability, and their influence on students' critical thinking. For instance, the performance of *ChatGPT* across physics tasks is highly variable, ranging from 15% to 80% accuracy, which may partly reflect its English-language training base (Revalde et al., 2025). *ChatGPT* performs well on conceptual physics questions but struggles with open-ended or calculation-heavy tasks, reflecting its strength in language rather than quantitative reasoning (Bender et al., 2021; Gregorcic and Pendrill, 2023; Revalde et al., 2025).

Students often lack the experience to critically assess such outputs, and may accept them without questioning them (Ding et al., 2023). In this context, encouraging students to examine responses carefully can become an important part of *ChatGPT's* educational use. When used thoughtfully, the tool may help develop critical thinking by prompting reflection on varying outputs. It may also serve as a starting point for discussions about the strengths and limitations of AI tools and their role in assessing the reliability of information (Bitzenbauer, 2023).

## 2.4 Learning theory integration

The Self-Determination Theory examines people's basic psychological needs that drive motivation as well as learning (Ryan and Deci, 2000). This theory distinguishes between two types of motivation: Intrinsic motivation is defined as the natural drive toward learning, skill development, curiosity and discovery, while extrinsic motivation means performing an activity to achieve a specific external goal (Ryan and Deci, 2000). Both types of motivation do not combine cumulatively (Deci, 1975).

Self-Determination Theory therefore highlights the significance of autonomy and competence in fostering intrinsic motivation for learning (Deci and Ryan, 2008). To effectively cultivate these conditions, it is essential to design educational experiences that not only support student motivation but also account for cognitive load. This refers to the demand placed on working memory by incoming information.

Cognitive load theory is an instructional framework grounded in understanding human cognitive architecture, specifically designed to address working memory constraints (Sweller, 2012). It seeks to enhance productive cognitive processing to facilitate knowledge acquisition. There are three different types of cognitive load, that can be aggregated (Sweller, 2012). Intrinsic cognitive load refers to the mental effort required due to the difficulty of the material being learned (Sweller, 2012). Extraneous cognitive load results from instructions that disregard working memory capacity and diverts mental resources away from building knowledge structures and germane cognitive load represents the cognitive demand generated by focused learning that leads to building knowledge (Sweller, 2012). Further, favorable emotional states that strengthen mental functions and reduce inherent complexity can generate additional working memory capacity that can be redirected to amplify learning experiences (Young et al., 2021).

Cognitive load theory emphasizes the importance of managing mental effort during learning, which aligns closely with Vygotsky's concept of the zone of proximal development (ZPD). This zone describes the gap between what learners can accomplish on their own and what they can achieve with guidance by a More Knowledge Other (MKO) (Rigopouli et al., 2025). Technology-enhanced approaches, including AI, function as scaffolding within the ZPD to help students reach beyond their independent capabilities (Rigopouli et al., 2025).

Generative AI like ChatGPT can act like an MKO and support learning processes by enabling learners to interact more thoroughly with the content (Wang and Fan, 2025). By guiding and supporting learners, extraneous cognitive load can be lowered and working memory resources released, potentially allowing for increased germane cognitive load (Sweller, 2012) and therefore better learning outcomes.

## 2.5 Emotional and motivational impacts

In addition to cognitive performance, several studies have explored how AI-based tools affect students' emotional and motivational experiences during learning. A study from Stanford found that while *ChatGPT* performs good on well-defined physics problems, it struggles with open-ended questions that require assumptions or contextual understanding (Wang et al., 2024). Fluent AI responses can create an illusion of understanding (Dahlkemper et al., 2023), while students' tendency to copy answers without reflection further limits critical engagement (Krupp et al., 2023a). This tendency can make them more likely to accept false or misleading content without questioning it (Kasneci et al., 2023; Kortemeyer, 2023), which in turn may reinforce misconceptions and negatively affect future learning (Ding et al., 2023). At the same time, students have rated AI as moderately important in physics education but viewed *ChatGPT* as a valuable tool for various applications (Bitzenbauer, 2023).

On the other hand, AI also can be beneficial in encouraging critical thinking and engagement. A recent study comparing moderated and unmoderated LLM use as well as simple internet searches suggests that the use of moderated LLMs can potentially promote critical thinking and enhance engagement strategies (Krupp et al., 2023b).

Possible positive motivational effects of an AI tool (Hanum Siregar et al., 2023) appear linked to the interactive and novel nature of the tools (Kuleto et al., 2021; Kestin et al., 2025; Wang and Fan, 2025). New technologies often attract attention to the tool itself rather than their educational function, which may result in superficial or unfocused use (Miguel-Alonso et al., 2024).

Turning to the impact on learning processes, studies indicate that students were able to learn effectively through *ChatGPT*, leading to improved understanding and higher achievement (Lo, 2023; Alarbi et al., 2024). AI learning tools have been reported to affect university students' learning experiences (Wen et al., 2024), increasing engagement and promoting positive attitudes toward learning (Gada and Chudasana, 2024). AI chatbots have been examined for their potential to support the development of practical skills in learning contexts. Research indicates that their use enhances cognitive abilities and supports skill development (Essel et al., 2024).

When used in educational settings, AI chatbots have been found to affect several aspects of student learning, including performance, motivation, interest, self-efficacy, and anxiety. However, factors like educational level and intervention duration moderate these effects (Wu and Yu, 2024). In higher education, where most studies on *ChatGPT* have been conducted (Lo, 2023), students generally view AI tools like *ChatGPT* as useful and engaging, though they also expect high levels of reliability, user-friendliness, and personalized support (Wen et al., 2024).

A comprehensive analysis of various studies on the impact of *ChatGPT* on education reveals that students' emotional engagement with a chatbot is primarily influenced by the perceived benefits, performance expectations and the quality of information output (Lo, 2023). Nevertheless, students utilizing generative AI demonstrate elevated cognitive as well as emotional involvement (Guo et al., 2025). Additional factors, such as enjoyment, self-directed regulation, ease of use and the alignment of outcomes with prior expectations, further contribute to engagement with the tool (Lo, 2023). Learners who view generative AI as beneficial for their studies display stronger positive emotional involvement, indicating that attitudes toward the technology significantly influence its impact (Guo et al., 2025). However, this enthusiasm with technology coexists with significant public skepticism (Vodafone Stiftung Deutschland gGmbH., 2023).

In summary, while AI tools show potential for enhancing engagement and offering personalized support, their impact on learning outcomes remains inconsistent, particularly in tasks requiring deeper conceptual understanding. This study takes up that challenge by comparing AI-supported learning with the more traditional tool *Microsoft Excel* in the context of physics experiments.

## 3 Methodology

To compare the educational impact of AI-assisted vs. *Excel*-based data analysis in physics, this study implemented a controlled intervention with student teachers analyzing pendulum experiment data. The intervention was implemented as a structured task in which participants evaluated two physics experiments, one involving a thread pendulum and the other a spring pendulum, using either the AI chatbot *ExperiMentor* or *Excel*.

The study followed a two-stage testing procedure to evaluate participants' experiences and skills. One group worked with *ExperiMentor*, a specialized conversational agent developed specifically for this study based on the GPT-4o model. *ExperiMentor* was implemented using the OpenAI GPT Creator and utilized between June and August 2024. It is not a publicly hosted tool apart from the OpenAI *ChatGPT GPT*-Store[1], nor integrated into an external platform. *ExperiMentor* was developed solely for research purposes without any commercial interest. To ensure transparency and reproducibility, the full system instructions and a screenshot of a sample interaction with *ExperiMentor* are provided in the Supplementary material.

*ExperiMentor* is a custom *GPT*-based chatbot that supports data analysis by automatically generating and executing *Python* code. Students received only the outputs (e.g., results, visualizations), with optional code previews, but no direct interaction with the code environment. *ExperiMentor's* behavior was governed by a system prompt designed around three core pillars:

First, contextual integrity and anonymity: the prompt enforces a strict authentication protocol. Before any pedagogical interaction begins, *ExperiMentor* is programmed to persistently request a generated identification code (based on demographic data) to link chat logs with pre-post survey data. The instructions explicitly forbid from proceeding without this code, ensuring data validity.

Second, pedagogical scaffolding and role definition: the system prompt defines *ExperiMentor's* role as a mentor or guide rather than a solver. It explicitly suppresses direct answer-giving behavior. Instead, the prompt instructs the model to guide students through data collection, analysis and interpretation. It promotes critical thinking by providing explanations, formulas, and hints only upon request or when errors are detected.

Third, interaction flow control: to manage cognitive load, the prompt enforces a step-by-step interaction model. After each analytical step, *ExperiMentor* is instructed to pause, display intermediate results, and ask the student *"What should we do next?"*. This design prevents the AI from hallucinating a full solution path and forces the student to maintain agency over the scientific process.

The prompts also included negative constraints to mitigate common LLM issues like long answers or over-helpfulness. These measures increased internal validity by keeping the pedagogical behavior and analytical guidance of *ExperiMentor* consistent throughout the intervention, regardless of individual student phrasing.

To evaluate the impact of the intervention, a randomized pre-post control group design was employed (Döring et al., 2016). Paired observations were used to assess change, comparing pre- and post-intervention responses within the same participants (Hedderich and Sachs, 2016). Both groups, one using *Excel*, the other *ExperiMentor*, completed identical tasks, allowing for direct comparison of outcomes.

### 3.1 Research questions

Despite the growing interest in AI tools, there is still no conclusive empirical evidence on their impact on learning processes. This emphasizes the urgent need for clear guidance on their role in educational practices, including their potential benefits, limitations, and challenges (Ständige Wissenschaftliche Kommission der Kultusministerkonferenz, 2024). As tools like *ChatGPT* are still at an early stage of adoption, ongoing research is crucial to understanding their effectiveness and ensuring their responsible integration into education (Farrokhnia et al., 2024). To help address this research gap, the present study investigates the following questions.

*Research Question 1 (RQ1):* how do students' results change from pre- to post-test within each group, and how do the learning gains compare between the *Excel* and AI groups? This research question combines both intragroup improvements and intergroup comparisons. It aims to analyze whether participants show measurable gains in their understanding of physical concepts and in their ability to apply data analysis methods after the

1 ExperiMentor can be accessed via: https://chatgpt.com/g/g-WZ4GsKZbZ-experimentor-studie (to interact with it, make up an identification code it will ask for).

intervention. It also explores whether the learning gains vary depending on the tool used.

*Research Question 2 (RQ2):* how do students' emotional responses differ when analyzing data using *Excel* vs. *ExperiMentor*? This question moves beyond cognitive learning outcomes and considers the emotional dimension of the learning experience. It explores whether the type of tool used influences students' emotional reactions, such as uncertainty, enjoyment, or frustration, while working on physics data analysis tasks. The goal is to develop a more holistic understanding of how digital tools affect engagement and motivation.

## 3.2 Procedure

The intervention, i.e., the instructional activity, was carried out by two different groups consisting of randomly assigned participants. Both groups evaluated one experiment on the thread pendulum and one on the spring pendulum using given data sets. Both groups were guided through the evaluation step by step with the help of identical tasks given on paper (see Supplementary material for details). The aim of the thread pendulum experiment was to determine the magnitude of the gravitational acceleration g. The spring pendulum experiment aimed to calculate the spring constant k.

The participating students selected seats at random from pre-prepared workstations, each pre-loaded with either *Excel* or *ExperiMentor*. The software configuration was hidden during selection to ensure unbiased assignment. The pre-test was carried out by the participants before they knew which group they belonged to. The task sheets were only made available after completion of the pre-test and removed again before the start of the post-test.

While *Excel* is not a pedagogical method, it was chosen as a control condition due to its widespread use in university-level data analysis. The goal was to provide a familiar, static environment that allowed for a contrast with the dynamic and responsive features of the AI-based chatbot. The *Excel* group worked without external resources to simulate a closed-book setting. To minimize potential cross-group influence, particularly given that both groups worked in the same physical space, participants were explicitly instructed not to discuss the tasks or digital tools during the session. The room layout was arranged to separate the groups and the intervention was monitored to ensure compliance with these instructions. Although minimal cross-group communication was possible, the room layout ensured working independence.

## 3.3 Material

The pre- and post-tests focused on evaluating participants' knowledge and skills in analyzing physical experiments. Emotional and motivational variables were assessed through structured survey instruments (see Supplementary material for details).

As Table 1 shows, the study followed a pre-post design consisting of five sections (Parts A to E), with a total of 31 items in the pre-test and 39 items in the post-test. Part A collected demographic information in the pre-test and group affiliation in the post-test.

Part B focused on emotional and motivational variables, using Likert-type items. In the pre-test, technology commitment (B1) was assessed using 7 validated items from the short scale for measuring technology commitment (Neyer et al., 2016), rated on a 5-point Likert scale. Participants also rated their interest and perceived ease of experimental evaluation (B2) using two 4-point Likert scales.

In the post-test, emotional responses were addressed in Section B3. Here, positive emotional learning experiences (items B3.1 and B3.2) captured participants' feelings of enjoyment and success, while negative emotional experiences (items B3.3 to B3.5) reflected emotional states such as uncertainty, stress, or frustration. Sections B4 and B5 examined participants' perception of the method and the perceived difficulty of concepts and topics, respectively. Section B6 focused on method effectiveness, which was subdivided into four constructs: Effectiveness of the method for understanding physics and supporting data analysis (items B6.1, B6.3), perceived learning gains regarding comprehension and progress (B6.2, B6.6), Motivation to engage with the experiment and evaluating data (B6.4, B6.5), and a comparison between the AI method and *Excel* (B6.7, B6.8).

Parts C to E assessed participants' conceptual understanding of thread and spring pendulum physics, as well as their data analysis competencies. The tasks in these sections consisted of objective, single-choice items designed to assess factual knowledge, conceptual reasoning, and the ability to interpret and analyze experimental data. These measures were used to evaluate actual learning gains between the pre- and post-tests.

The group working with *Excel* received an *Excel* spreadsheet in which the respective measured values of the experiment were already entered. They could edit the table freely and were not subject to any restrictions, except that they were not allowed to search for help outside of *Excel* and the printed task sheets (see Supplementary material). The group using the AI tool *ExperiMentor* completed the tasks within the chat environment of the custom GPT *ExperiMentor*. They also received an *Excel* spreadsheet containing the measurement results. They had the option of transferring the data manually, copying it or uploading the entire table to the chat.

## 3.4 Data collection and analysis

Data were recorded using a digital survey tool and analyzed in *R Studio* (version 4.3.3) (Hanum Siregar et al., 2023). Descriptive parameters included mean (M), median (MDN), and standard deviation (SD (Miguel-Alonso et al., 2024). The Shapiro-Wilk test was used to check distribution assumptions (Miguel-Alonso et al., 2024) because of its sensitivity to deviations from normality (Kortemeyer, 2023).

### 3.4.1 Performance data

To address the first research question, learning gains within groups between pre and post were examined. Parts C, D, and E of the questionnaire were analyzed first, followed by the three subject areas thread pendulum, spring pendulum as well as evaluation methods and finally individual questions. The *Shapiro-Wilk test* indicated non-normal distribution throughout. A two-sided *Wilcoxon signed-rank test* ($\alpha \leq 0.05$) was used to check for significant directional changes between the groups and the pre- and post-test (Wollschläger, 2014). To assess effect sizes the rank-biserial correlation was used (Kerby, 2014).

TABLE 1 Structure of the pre- and post-test instruments, including item count and response format by section and focus area.

| Section | Focus | # Items | Scale |
|---|---|---|---|
| Part A (Pre/Post) | Demographics | 4 Pre 1 Post | |
| Part B (Pre/Post) | Emotional & motivational Variables | 9 Pre 20 Post | Likert Scales |
| B1 (Pre) | Technology commitment | 7 | 5-point Likert |
| B2 (Pre) | Interest & perceived ease of experimental evaluation | 2 | 4-point Likert |
| B3 (Post) | Emotional learning experience | 5 | 4-point Likert |
| B4 (Post) | Perception of method | 3 | 4-point Likert |
| B5 (Post) | Difficulty of concepts & topic | 4 | 4-point Likert |
| B6 (Post) | Method Effectiveness | 8 | 4-point Likert |
| Part C (Pre/Post) | Thread pendulum physics | 6 | Nominal (Single Choice) |
| Part D (Pre/Post) | Spring pendulum physics | 6 | Nominal (Single Choice) |
| Part E (Pre/Post) | Data analysis competencies | 6 | Nominal (Single Choice) |

An *ANCOVA* was used examining the intervention's impact while controlling for pre-test scores (Bortz and Schuster, 2010; Wollschläger, 2014). Despite non-normal distribution, ANCOVA is robust against such violations (Glass et al., 1972; Bortz and Schuster, 2010; Schmider et al., 2010), especially with equal group sizes ($n = 25$ each) (Bortz and Schuster, 2010). To evaluate the *ANCOVA* assumption of homogeneous regression slopes, it was tested whether the covariate interacted with the experimental factors. The interaction between Group and Pre-Test was not significant in any model, indicating that the basic assumption of parallel slopes across groups was met. However, the higher-order interactions involving the covariate, particularly Group × Measurement × Pre were significant in all models (see Table 6). This indicates that the relationship between the covariate and the dependent variable varied across measurement points and groups, meaning that full homogeneity of regression slopes was not strictly satisfied. Consequently, the ANCOVA results should be interpreted with caution.

The analysis began with the total sum of items to compare general learning gains, with pre-test scores as covariates. The three main categories were then analyzed separately, followed by individual items within each category. The *Hake Index g* was calculated for combined and separate group data. This metric measures average normalized learning growth (Hake, 1998) as the proportion of actual to maximum possible improvement (McKagan et al., 2022). The gain of averages method was used to evaluate the effectiveness of the learning interventions (McKagan et al., 2022). The index shows minimal correlation with pre-test scores, indicating independence from prior knowledge. In contrast, the average post-test score and the average gain are less suitable for comparing the course success rate across different groups (Hake, 2002; Coletta and Steinert, 2020). Values for $g$ range from 0 (no gain) to 1 (maximum gain; Hake, 2002).

### 3.4.2 Emotional-motivational data

Descriptive statistics and Shapiro-Wilk tests were conducted for all questions. Internal reliability was evaluated using *Cronbach's* $\alpha$, which represents the average correlation among items (Döring et al., 2016) and the proportion of test variance that can be attributed to shared factors across the items (Cronbach, 1951). The *Wilcoxon rank-sum test* measured differences between groups, as it does not require normal distribution (Wollschläger, 2014), which was not given after analyzing the results of the *Shapiro-Wilk test.*

## 3.5 Sample

The $n = 50$ participants were student teachers from the University of Cologne: 33 male, 17 female, aged 19–37 years ($M = 24.04$). At the time of the study, 35 were in a Bachelor of Arts program and 15 in a Master of Education program. The Excel group ($n = 25$) had an average age of $M = 24.36$ years, with 17 Bachelor and 8 Master students. The AI group ($n = 25$) had an average age of $M = 23.72$ years, with 18 Bachelor and 7 Master students. All participants were training to become teachers for various school types.

# 4 Results

This section presents the results of the experimental study in accordance with the two research questions. Section A reports learning-related outcomes based on participants' pre- and post-test performance, while Section B addresses emotional and motivational outcomes assessed through survey responses.

## 4.1 Learning outcomes (RQ1)

### 4.1.1 Descriptive statistics

Descriptive statistics were calculated to provide an overview of participants' performance before and after the intervention (see Tables 2, 3 and Figure 1). For both groups, *mean scores (M)*, *medians (MDN)*, and *standard deviations (SD)* were determined for the total score and the three domains: thread pendulum (TP), spring pendulum (SP), and data evaluation (DE).

Items within the thread pendulum domain focused on conceptual understanding and application of the physical principles of a simple thread pendulum, including the influence of length, gravitational acceleration and probing mass independence as well as scenarios involving altered gravity or acceleration. The spring pendulum domain included items regarding the factors determining the period of a spring pendulum, such as dependence on mass, gravitational acceleration and the spring constant. Items within the data evaluation (DE) domain required knowledge on the interpretation of graphical data, application of linear regression and understanding of statistical metrics. Overall, both groups tended to perform better after the intervention, with increases in scores seen across the total results and in most of the tested areas. Score variability remained stable overall, with a slight narrowing in the AI group's post-test data.

TABLE 2 Descriptive data for the excel group.

| Item | M | | SD | | MDN | |
|---|---|---|---|---|---|---|
| | Pre | Post | Pre | Post | Pre | Post |
| Total | 7.72 | 8.96 | 3.27 | 3.51 | 7 | 9 |
| TP | 3.32 | 3.16 | 1.80 | 1.77 | 3 | 3 |
| SP | 1.56 | 1.96 | 1.39 | 1.54 | 1 | 2 |
| DE | 2.84 | 3.84 | 1.31 | 1.28 | 3 | 4 |

TABLE 3 Descriptive data for the AI group.

| Item | M | | SD | | MDN | |
|---|---|---|---|---|---|---|
| | Pre | Post | Pre | Post | Pre | Post |
| Total | 7.64 | 9.92 | 3.43 | 2.96 | 7 | 9 |
| TP | 3.24 | 3.72 | 1.45 | 1.40 | 3 | 4 |
| SP | 1.40 | 2.20 | 1.29 | 1.29 | 1 | 2 |
| DE | 3.00 | 4.00 | 1.61 | 1.22 | 3 | 4 |

### 4.1.2 Intragroup learning gains

To assess the learning gains within each group, the results of the pre- and post-tests were analyzed separately for the *Excel* and AI groups using the *Wilcoxon signed-rank test*. Effect sizes (*rank biserial correlation*) indicate the magnitude of learning gains across domains. The effect sizes for the rank-biserial correlation are categorized as very small (< |0.10|), small (|0.10| to |0.29|), moderate (|0.30| to |0.49|), or large (≥ |0.50|; López-Martín and Ardura, 2023).

The results of each group are presented in Tables 4, 5. The subsequent analysis applies the same statistical approach to the second group, enabling a consistent assessment of learning progress across domains.

TABLE 4 Wilcoxon test and effect sizes excel group.

| Item (excel) | *p*-value | Rank biserial correlation |
|---|---|---|
| Total | 0.04 | −0.54 |
| TP | 0.60 | |
| SP | 0.32 | |
| DE | <0.01 | −0.78 |

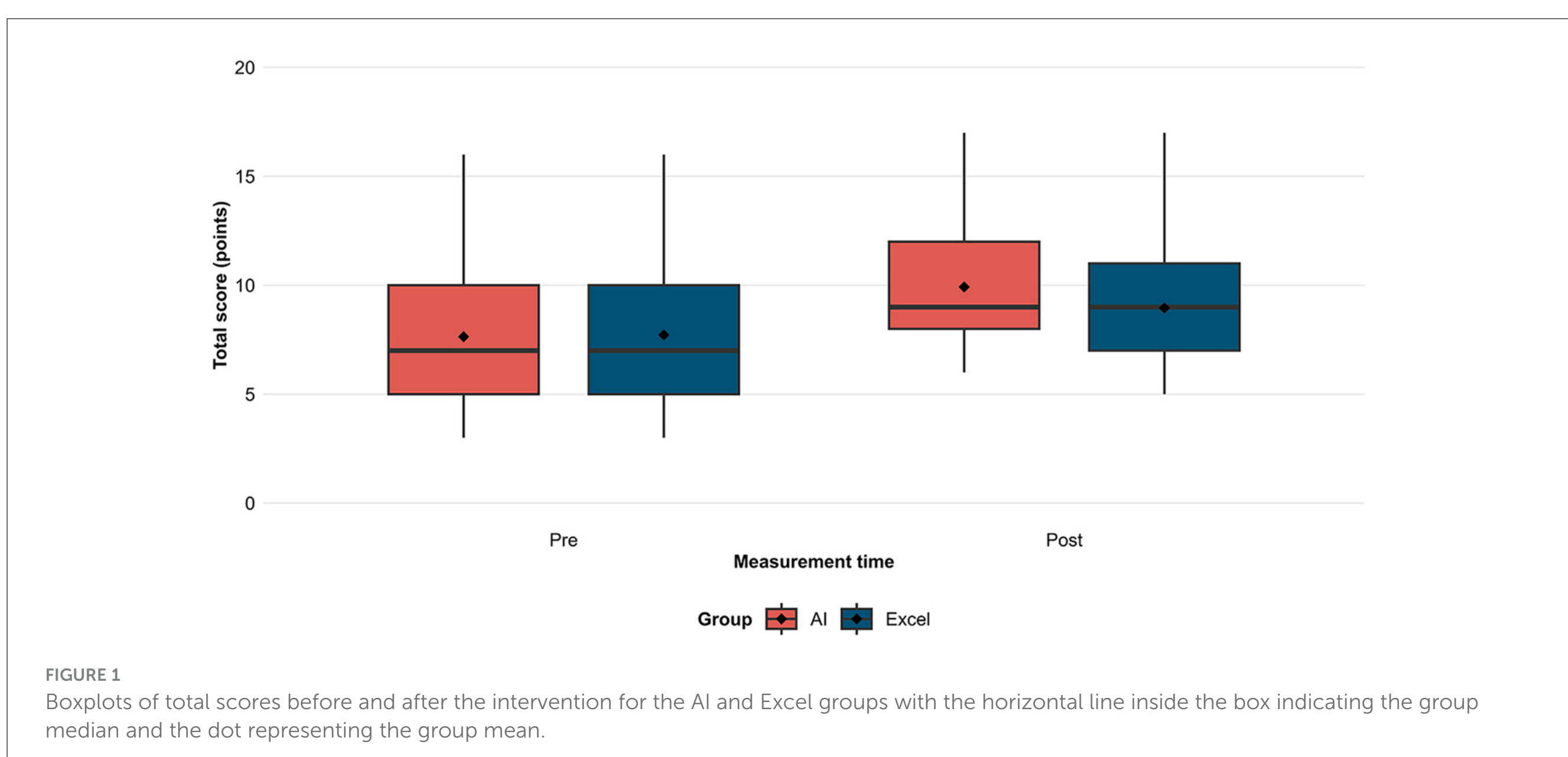

FIGURE 1
Boxplots of total scores before and after the intervention for the AI and Excel groups with the horizontal line inside the box indicating the group median and the dot representing the group mean.

Both groups showed significant learning gains from pre- to post-test, though the magnitude and distribution of these gains differed by domain. In the *Excel* group, only the data evaluation domain showed a statistically significant improvement, with a large effect size. The other two domains showed no significant change. By contrast, the AI group exhibited significant gains in both the spring pendulum and data evaluation domains, with very large effect sizes. The thread pendulum domain reached a borderline significance threshold, suggesting a possible effect that warrants further investigation. Overall, the total scores improved significantly in both groups, with the AI group displaying a notably larger effect.

### 4.1.3 Intergroup learning gains

To investigate differences in learning gains between the two groups, an *ANCOVA* was conducted (see Table 6). The dependent variable was the post-test performance, the independent variable was the group assignment, and the pre-test scores served as the covariate. This design allows for a more accurate comparison of group outcomes by adjusting for individual differences in performance. *ANCOVA* was used despite non-normality (Glass et al., 1972; Bortz and Schuster, 2010). Partial $\eta^2$ values showed that the group factor accounted for a negligible amount of the variance in all models (see Table 6). In contrast, the measurement factor showed small to moderate effects, particularly for the total score and the DE items.

In addition to that, the *Hake Index (g)* was calculated (see Table 6) to quantify normalized learning gains. This measure provides a standardized way to compare relative learning progress regardless of initial performance levels, where values below 0,3 indicate low gains and values near 0 or negative suggest minimal or no improvement (Neyer et al., 2016).

No statistically significant differences were found between the AI and *Excel* groups at the overall level or in any of the specific domains. The total *Hake Index* value of $g = 0{,}02$ also indicates only marginal learning progress when combining both groups.

While *ANCOVA* offers adjusted comparisons between groups, the *Hake Index* complements this by summarizing the magnitude of within-group progress relative to potential gains. To better understand group-specific improvements, the Hake Index was also calculated separately for each condition and is presented in Table 7.

Overall, the results from both the *ANCOVA* and *Hake Index* calculations indicate that no statistically significant differences were observed between the groups across the measured domains. The *Hake Index* values indicated minimal learning gains in both groups.

TABLE 5 Wilcoxon test and effect sizes AI group.

| Item (AI) | *p*-value | Rank biserial correlation |
|---|---|---|
| Total | <0.01 | −1 |
| TP | 0.05[a] | −0.53 |
| SP | < 0.01 | −0.8 |
| DE | < 0.01 | −0.86 |

[a]Borderline significant.

TABLE 7 Hake indices per group.

| Item | Hake *g* AI | Hake *g* excel |
|---|---|---|
| Total | 0.02 | 0.01 |
| TP | 0.01 | −0.004 |
| SP | 0.02 | 0.01 |
| DE | 0.02 | 0.02 |

TABLE 6 ANCOVA results and Hake Index summary.

| Item | *p*-value group × measurement | Slope homogenity pre × group × measurement | Partial $\eta^2$ | | Hake *g* (total) |
|---|---|---|---|---|---|
| Total | 0.39 | <0.01 | Group | <0.01 | 0.02 |
| | | | Measurement | 0.08 | |
| | | | Group×Measurement | <0.01 | |
| TP | 0.30 | < 0.01 | Group | <0.01 | < 0.01 |
| | | | Measurement | <0.01 | |
| | | | Group × Measurement | 0.01 | |
| SP | 0.45 | <0.01 | Group | <0.01 | 0.01 |
| | | | Measurement | 0.05 | |
| | | | Group × Measurement | <0.01 | |
| DE | 1 | <0.01 | Group | <0.01 | 0.02 |
| | | | Measurement | 0.13 | |
| | | | Group × Measurement | 0 | |

## 4.2 Emotional and motivational outcomes (RQ2)

### 4.2.1 Pre-intervention differences in emotional-motivational attitudes

Before the intervention, participants in both groups completed survey items designed to capture their emotional-motivational attitudes toward technology and experimental data analysis. Section B1 included seven previously validated items on technology commitment (Neyer et al., 2016), while Section B2 contained two items assessing interest in and perceived ease of evaluating experiments. Given the small number of items in B2 and the prior validation of B1, no reliability analysis was performed for the pre-test.

Due to non-normality, group comparisons were conducted using *Wilcoxon rank-sum tests* (see Table 8). Across all items, no statistically significant differences were found between the *Excel* and AI groups. This suggests that the groups started the intervention with comparable attitudes toward both technology and the task of evaluating experimental data.

### 4.2.2 Post-intervention differences in emotional-motivational attitudes

After completing the experimental tasks, participants responded to a broader set of questions grouped into eight constructs reflecting their emotional and motivational experience (see II. B. in the Supplementary material). These included categories such as enjoyment, frustration, perceived effectiveness, and motivation. Internal consistency of the constructs was evaluated using *Cronbach's* $\alpha$. Most constructs reached acceptable to good reliability, with only two falling slightly below the conventional threshold. The calculated $\alpha$ as well as mean scores, significant values p and effect sizes are presented in Table 9.

Violin plots revealed clear differences in responses (see Figure 2). Responses in the AI group clustered toward the higher end of the scale, particularly in constructs such as *Positive Emotional Learning Experience*, *Method Effectiveness*, and *Motivation*. In contrast, the *Excel* group exhibited broader distributions with more lower-end responses, especially in Negative *Emotional Learning Experience* and *Method Comparison*.

TABLE 8 Wilcoxon rank-sum test results for pre-test items.

| Item | Formulated | *p*-value |
|---|---|---|
| B1.1 | I am very curious about technical innovations. | 0.41 |
| B1.2 | I am often afraid of failing when dealing with modern technology. | 0.36 |
| B1.3 | If I had the opportunity, I would use technical products much more frequently than I currently do. | 0.85 |
| B1.4 | I quickly develop a liking for technical innovations. | 0.39 |
| B1.5 | I find dealing with new technology difficult – I usually can't manage it. | 0.71 |
| B1.6 | Dealing with technical innovations mostly overwhelms me. | 0.99 |
| B1.7 | Whether I succeed in using modern technology depends mainly on me. | 0.17 |
| B2.1 | I find it easy to evaluate experiments. | 0.57 |
| B2.2 | I find evaluating experiments interesting. | 0.87 |

TABLE 9 Group differences in emotional-motivational constructs: mean scores, significant values (*p*), effect sizes (rank biserial correlation) and internal consistency (Cronbach's $\alpha$).

| Construct | Items | $\alpha$ | Mean | | *p*-value | Effect size |
|---|---|---|---|---|---|---|
| | | | AI | Excel | | |
| Positive emotional Learning experience | B3.1 B3.2 | 0.77 | 2.26 | 1.74 | 0.01* | −0.392 |
| Negative emotional Learning experience | B3.3 B3.4 B3.5 | 0.82 | 0.64 | 1.56 | <0.01* | 0.717 |
| Perception of methods | B4.1 B4.2 B4.3 | 0.73 | 2.44 | 1.68 | <0.01* | −0.65 |
| Challenges | B5.1 B5.2 B5.3 B5.4 | 0.80 | 0.93 | 1.13 | 0.33° | 0.294°° |
| Method effectiveness | B6.1 B6.3 | 0.85 | 2.52 | 1.88 | <0.01* | −0.459 |
| Perceived learning gains | B6.2 B6.6 | 0.68 | 1.38 | 1.26 | 0.66 | −0.07 |
| Motivation | B6.4 B6.5 | 0.62 | 2.28 | 1.56 | < 0.01* | −0.0584 |
| Method comparison | B6.7 B6.8 | 0.81 | 1.04 | 1.89 | < 0.01* | 0.646 |

°t-test.
°°Cohen's d.
*Significant.

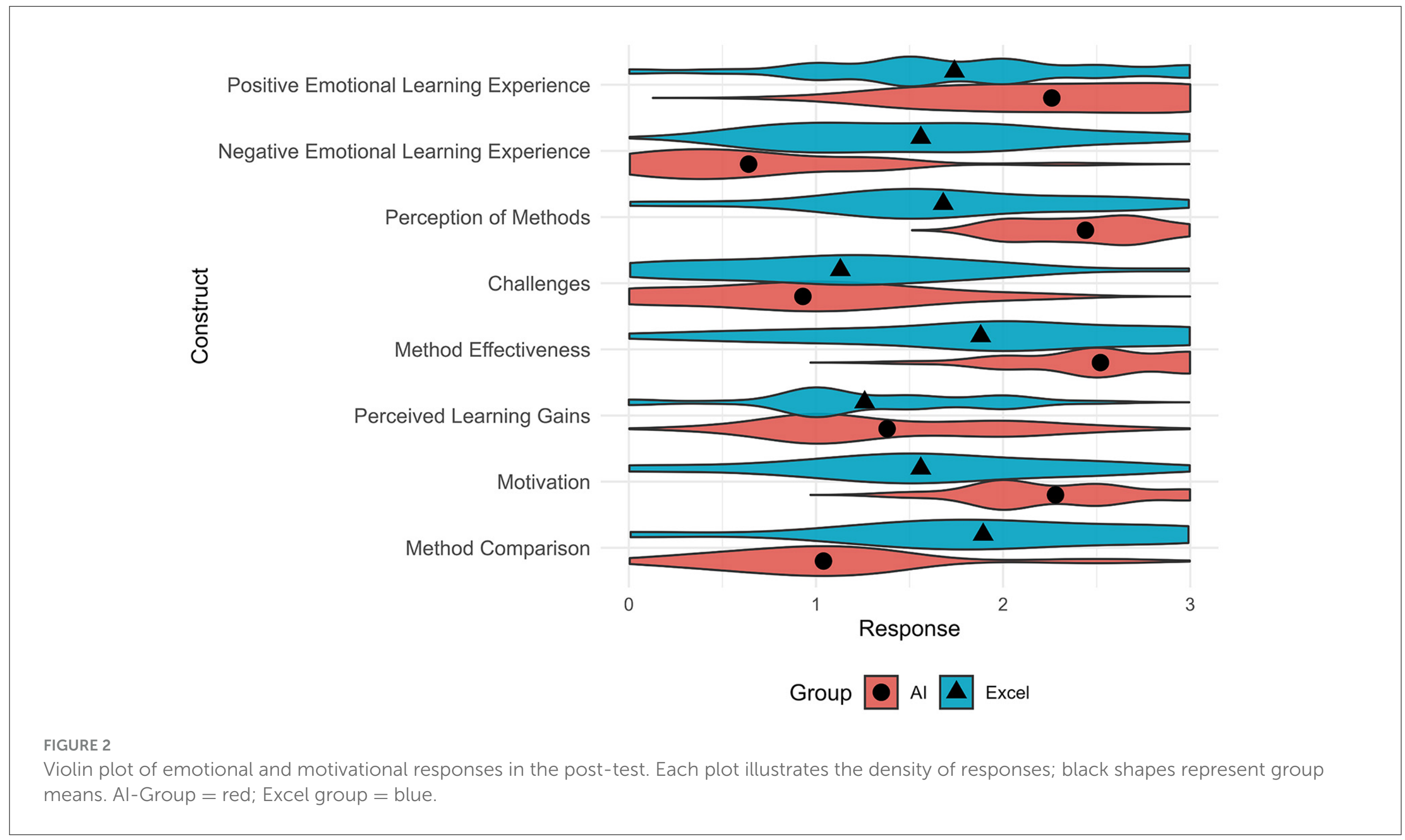

FIGURE 2
Violin plot of emotional and motivational responses in the post-test. Each plot illustrates the density of responses; black shapes represent group means. AI-Group = red; Excel group = blue.

Overall, the data indicates clear differences in emotional and motivational outcomes between the groups. The constructs *Positive Emotional Learning Experience, Perception of Methods, Method Effectiveness, Motivation*, and *Method Comparison* showed statistically significant differences between the AI and *Excel* groups, with medium to large effect sizes. In contrast, *Negative Emotional Learning Experience* was significantly lower in the AI group. Other constructs such as *Challenges* and *Perceived Learning Gains* did not show significant differences. These results reflect distinct patterns in how students experienced and evaluated the respective learning tools.

# 5 Discussion

This section interprets the study's findings considering the research questions and the existing literature. It considers both the learning-related outcomes and the emotional-motivational dimensions of the intervention, before reflecting on broader methodological implications and practical considerations.

## 5.1 Interpretation of learning outcomes (RQ1)

The first research question focused on the extent to which students benefited from the intervention in terms of content learning. This section discusses both within-group improvements and between-group comparisons.

### 5.1.1 Intragroup learning gains

The Wilcoxon analyses indicate that both tools supported some degree of learning, although the size and distribution of these gains varied by domain. The *Excel* group showed a moderate improvement overall, with the most notable gain occurring in the domain of data evaluation. Improvements in the *Excel* group likely reflect familiarity with the tool and the structured task design, rather than the tool's pedagogical value. No statistically significant gains emerged for the thread or spring pendulum tasks, which suggests that the structured Excel workflow may reinforce procedural aspects of data handling but offers limited support for more conceptually demanding domains. From a Cognitive Load Theory perspective *Excel's* technical interface with many interactive elements may have imposed excessive extraneous cognitive load through its manual data manipulation requirements, making the content appear complex and challenging (Sweller, 2012). Lacking adaptive feedback, students likely expended their limited working memory on extraneous load rather than germane load. Consequently, cognitive resources that could have been devoted to germane processing and conceptual understanding were instead consumed by managing operational demands. This is consistent with the relatively low Hake values in this group. Unlike *ExperiMentor*, *Excel* provided no responsive feedback, which may have contributed to lower motivation and higher frustration.

In contrast, the AI group showed significant gains across most domains, particularly in data evaluation and spring pendulum tasks. The effect sizes observed suggest that *ExperiMentor* was effective in supporting students' learning progress. Its capability to generate explanations, visualizations, and real-time feedback likely contributed to this outcome. This finding can be interpreted through Vygotsky's ZPD, where *ExperiMentor* functioned as a MKO by providing scaffolded support tailored to students' current knowledge levels. The stronger gains in the AI group can therefore be understood as the result of learning activities that remained within students' ZPD while still extending their independent performance. The AI's capacity to offer immediate, contextualized guidance may have enabled learners to navigate tasks beyond their independent capabilities, effectively working within their ZPD. By managing cognitive load and providing adaptive scaffolding, the AI tool could have reduced extraneous demands while supporting germane cognitive processes necessary for conceptual development (Ratniyom et al., 2025). This aligns with findings from other studies that emphasize the potential of AI tools to enhance engagement, guide reasoning, and provide tailored guidance in complex tasks (Gada and Chudasana, 2024; Küchemann et al., 2024; Wen et al., 2024). The Hake indices for this group were also small in absolute terms.

Limited gains in the thread pendulum task suggest that neither tool provided sufficient conceptual scaffolding for more abstract problem domains. As prior research has shown, *ChatGPT* tends to perform better on well-structured, theory-based prompts than on tasks requiring mathematical modeling or nuanced physical interpretation (Wang et al., 2024; Revalde et al., 2025). This domain-specific limitation might explain the absence of a stronger effect in this area, despite the learning gains in the AI group. The data evaluation tasks involve more procedural elements, where progress may appear limited when initial performance is already moderate. Conversely, the thread pendulum and spring pendulum require abstract conceptual reasoning, and low normalized gains likely reflect the inherent difficulty of this domain rather than the ineffectiveness of either tool.

The findings reflect the differentiated potential of each tool. While *Excel* remains a reliable tool for basic data processing, it appears to offer limited support for learning gains in conceptual domains without additional instructional input. In contrast, the AI-assisted approach demonstrates potential for scaffolding more complex reasoning and analytical thinking. Other studies have shown the practicality and success of AI instructors in academic environments (Kestin et al., 2025). This further resonates with previous findings on the motivational and cognitive benefits of intelligent tutoring systems (Liang et al., 2023; Lo, 2023). However, these benefits were not uniform across all content areas, indicating that AI tools should not be viewed as a universal solution but rather as one component within a broader instructional strategy (Chounta et al., 2022).

### 5.1.2 Intergroup learning gains

While both groups improved from pre- to post-test, the *ANCOVA* results indicated no statistically significant differences in learning gains between the AI-assisted and *Excel*-based approaches. Across all learning domains, the group by measurement interaction was non-significant and the associated partial $\eta^2$ values were negligible, showing that the choice of tool did not meaningfully alter learning progress when initial performance was controlled. The covariate explained a substantial proportion of the variance in post-test scores, consistent with the well-established link between prior knowledge and later achievement. Assumption checks indicated that the relationship between pre- and post-test performance varied slightly across measurement points, suggesting that regression slopes were not fully homogeneous and that adjusted means should be interpreted with some caution. Nevertheless, the overall pattern remained clear: after controlling for baseline differences, both tools produced comparable learning outcomes.

Similarly, the *Hake Index* values were generally low across both groups. This does not necessarily indicate that students made negligible progress. Rather, they reflect the domain-specific nature of the tasks. Within this context, the AI tool's broader intragroup improvements and higher effect sizes reflect enhanced support for reasoning processes without producing measurable group-level advantages in the statistical analysis. This is consistent with the *ANCOVA* findings, where the tool variable explained only a minimal fraction of the variance and further with prior research noting that AI tools like *ChatGPT* tend to perform more reliably on conceptual or theory-based problems than on open-ended or quantitative tasks (Wang et al., 2024; Revalde et al., 2025).

Despite higher emotional engagement, the AI tool did not lead to significantly better performance outcomes compared to *Excel* under structured conditions. This discrepancy suggests a form of affective-cognitive dissonance, where learners feel more supported and confident without necessarily achieving superior objective learning gains. Because AI smooths the learning path by providing immediate answers and structural support, students may perceive the content as easier and themselves as more competent than their independent performance warrants. The tool effectively bridges the gap between the learner's ability and the task requirements, potentially masking the cognitive effort required to truly master the material. This aligns with earlier research that points to the context-dependence of AI tool effectiveness. While some studies report that tools like *ChatGPT* can outperform traditional platforms like *Excel* on specific tasks (Halaweh, 2023), other research emphasizes that these advantages often depend on how AI is embedded into the learning environment, the nature of the task, and the learner's ability to engage critically with the tool's output (Bitzenbauer, 2023; Gada and Chudasana, 2024).

The structured instructional design may have compensated for differences in tool affordances by providing explicit scaffolding that managed cognitive load for both groups. While *ExperiMentor* acted as an MKO within students' ZPD, the *Excel* group received comparable support through carefully designed task structures. In both cases, learners were provided with scaffolding that helped them operate near the upper boundary of their ZPD, which may explain the absence of pronounced between group differences in performance. This suggests that the presence of scaffolding, whether technology-based or instructionally embedded, can be more influential than the specific tool used.

Overall, these results show that AI tools do not automatically lead to superior learning outcomes compared to traditional methods. Instead, the effectiveness of any tool depends on how it is implemented. Similar to the findings of Halaweh (2023), this study found a not significant but noticeable performance advantage in the AI group. Both tools were embedded within a guided setting, which may explain the absence of major differences in learning outcomes.

## 5.2 Interpretation of emotional and motivational outcomes (RQ2)

Understanding students' attitudes regarding the used tools is key to evaluating its broader educational impact. This section reflects on participants' emotional and motivational responses, both before and after the tasks.

### 5.2.1 Pre-intervention attitudes

Before the intervention began, students in both groups held similar attitudes toward working with digital tools and evaluating experimental data. This was reflected in nearly identical mean scores across all items and negligible effect sizes. These findings indicate that neither group had an emotional or motivational advantage at the outset.

This comparability suggests that the emotional and motivational differences observed later are less likely to stem from pre-existing preferences or dispositions. Rather, they appear to be linked to the experience of working with the respective tools during the task itself.

### 5.2.2 Post-intervention attitudes

After the intervention, clear distinctions, as shown in Figure 2 emerged in how participants experienced the two tools. Learners in the AI group reported significantly more positive emotional learning experiences, including greater enjoyment and a stronger sense of success. These differences point to a learning environment that better supports basic psychological needs. The emotional advantages of the AI tool suggest that its interactivity and responsiveness fostered engagement and reduced negative emotional learning experiences like frustration. *ExperiMentor* appears to have better satisfied students' basic psychological needs for autonomy, relatedness and competence (Deci and Ryan, 2008), allowing more positive emotional learning experiences the students through personalized, responsive interactions (Chiu, 2025), adaptive feedback and conversational cooperative-seeming engagement (Ratniyom et al., 2025). Environments that support these needs promote better learning compared to environments that blocks these needs (Ryan and Deci, 2000). The AI tool's ability to provide individualized support likely enhanced students' sense of competence while its interactive nature fostered a perception of cooperation, even in a digital environment. As visualized in Figure 2, the distributional differences between groups further highlight these effects, particularly in constructs such as motivation and emotional experience.

These findings align with earlier studies suggesting that AI tools, especially those that simulate human-like interactions such as chatbots, can foster emotional engagement and reduce anxiety through scaffolding and adaptive feedback (Popenici and Kerr, 2017; Liang et al., 2023; Farrokhnia et al., 2024). At the same time, these outcomes should be interpreted with care. The emotional and motivational advantages may partly reflect novelty effects or the perception of receiving personalized assistance rather than a demonstrable educational benefit (Long et al., 2024).

However, attributing these differences solely to novelty risks oversimplification. It is critical to reflect on the role of feedback quality as a distinct design variable. The experimental setup compared a tool offering responsive scaffolding through *ExperiMentor* against *Excel*, offering no adaptive feedback. Consequently, the observed motivational gap may stem less from the AI itself and more from the discrepancy in interactivity. The static nature of the Excel condition represents a lack of feedback, whereas the AI condition provided continuous, conversational support. The positive outcomes in the AI group likely reflect the benefits of adaptive feedback mechanisms, which could theoretically be implemented via non-AI means, rather than the unique properties of AI alone.

The *Excel* group reported higher levels of negative emotional responses such as stress and frustration, emotional reactions can be seen as a consequence of elevated extraneous load. This contrast was associated with a very large effect size, reinforcing the interpretation that the AI-supported environment may have offered greater emotional comfort. Again, the static nature of *Excel* may have failed to manage extraneous cognitive load effectively, forcing students to simultaneously handle technical operations and conceptual understanding, contributing to increased frustration and diminished emotional engagement. The AI tool's ability to respond flexibly to user input and offer contextual hints may have provided users with a sense of support, thereby reducing uncertainty during problem-solving, a benefit noted in recent research on AI-supported learning (Liang et al., 2023; Gada and Chudasana, 2024; Mustofa et al., 2024; Tong et al., 2024).

Students' perception of the method also differed significantly between the groups. Participants rated the AI-based environment as more interesting, intuitive, and useful for future applications, with a very large effect size, suggesting a strong difference in tool preference. In SDT terms, this preference can be interpreted as a sign that students experienced the AI environment as more autonomy supportive and competence enhancing than the *Excel* setting. This reflects findings in the literature that associate AI tools with a higher degree of user satisfaction and perceived usefulness (Lo, 2023). Similarly, participants perceived the AI-supported method as more effective, particularly in helping them navigate complex or unfamiliar tasks, an interpretation supported by research suggesting that AI tools can strengthen learners' self-perceived performance (Wu and Yu, 2024).

Despite differences in engagement, both groups perceived the tasks as equally challenging, indicating that tool preference did not stem from task difficulty. This suggests that the more positive experience in the AI group was not due to an easier task, but perhaps due to the type and quality of feedback received.

While the two groups reported comparable perceived learning gains, motivation levels were notably higher in the AI group. This difference was also reflected in a large effect size highlighting a meaningful motivational advantage associated with the AI tool. This finding is consistent with research indicating that AI tools can increase learner motivation through interactivity and responsiveness (Hanum Siregar et al., 2023; Liang et al., 2023; Wu and Yu, 2024) and is consistent with Self Determination Theory, which attributes higher quality motivation to environments that satisfy learners' needs for autonomy, competence and relatedness (Deci and Ryan, 2008). The motivational advantage observed aligns with Self-Determination Theory's emphasis on intrinsic motivation emerging from satisfaction of basic psychological needs (Deci and Ryan, 2008). The AI tool's personalized responses likely enhanced perceived autonomy by allowing students to explore at their own pace, while its supportive feedback fostered competence perceptions. Furthermore, as an MKO, the AI provided scaffolding within students' ZPD. Similarly, through creating a climate of affective encouragement and acknowledgment, another study showed learners perceived themselves as capable of achieving their highest capabilities (La Ossa et al., 2024). Nevertheless, the disconnect between this high motivation and the comparable test scores reinforces the presence of an affective-cognitive dissonance. Students experienced a greater sense of competence and autonomy facilitated by *ExperiMentor*, even if this did not translate into superior quantitative performance.

Finally, responses to the Method Comparison construct showed a strong preference for the AI tool even among *Excel* users. The students across both groups saw clear advantages in the AI environment and believed it could have helped them complete the task more effectively, this could be by having reduced manual calculation effort or possible time savings by not having to manually navigate in the spreadsheet of Excel. Further, the additional feedback support and guidance embedded in *ExperiMentor*, which the *Excel* group lacked could be a reason for this tool preference. Lastly, since the AI handled the calculations, preparations and presentation, the extraneous cognitive load of the students might be reduced, which may lead to focus on the conceptual understanding and therefore preferring the AI tool. This redistribution of cognitive resources is exactly what Cognitive Load Theory describes as a shift from extraneous toward germane load. This cognitive redistribution, combined with the AI's role as an adaptive MKO operating within learners' ZPD, may explain the consistent preference across both groups.

The patterns displayed in Figure 2 confirm the differences identified in the quantitative analysis. The more concentrated distributions in the AI group suggest a consistently more positive experience, particularly regarding emotional support and perceived tool effectiveness. The *Excel* group showed greater variation, with several constructs demonstrating broader spreads toward lower ratings. This reinforces the interpretation that the AI-assisted environment offered a more uniformly positive experience.

## 5.3 Methodological reflections

This study explored the effects of AI-assisted vs. traditional data analysis using a randomized pre-post design in a real educational setting. It has limitations in duration, design, and sample characteristics. The intervention was relatively short, allowing only limited time for participants to fully engage with the tools. Consequently, the study may have captured only early-stage effects, rather than sustained learning trajectories. The pre- and post-tests focused on factual knowledge and practical application. However, closed-ended tasks may have missed complex learning forms such as conceptual change or skill transfer. The *Hake Index* and effect sizes provided useful information about average learning gains, but averages can mask individual differences. This study compared two different learning environments, not just two tools. The *Excel* group worked with a spreadsheet without feedback beyond written instructions. The AI group interacted with *ExperiMentor*, which responded to input, gave hints, offered feedback, and visualized results. This difference in interactivity shaped both performance and the learning experience. Although groups were seated separately and instructed not to interact, the shared physical space represents a potential source of cross-group influence. While incidental communication is unlikely to have substantially affected outcomes, it cannot be ruled out entirely.

## 5.4 Implications and limitations

The findings offer insights into the use of AI-assisted tools like *ExperiMentor* in physics education. However, several limitations must be considered when interpreting the results and assessing their broader relevance.

The relatively small, homogeneous sample of 50 student teachers from a single university restricts generalizability. The participants shared institutional context, prior knowledge, and cultural background limits the ability to draw conclusions about how *ExperiMentor* might function across diverse learning environments, educational systems, or disciplinary contexts beyond physics.

The limited impact observed in this study can be attributed to its single-session design. Research demonstrated that ChatGPT's influence on academic achievement becomes minimal with short-term applications under 1 week (Wang and Fan, 2025). Therefore, the short intervention duration limited deep tool exploration and made it difficult to evaluate lasting learning gains or long-term motivational effects. The single-session design offers limited insight into longitudinal outcomes or transfer effects to real classroom settings, where teachers would need to apply learned skills repeatedly and adapt them to varying student needs. Furthermore, the intervention relied on pre- and post-performance comparisons and self-reported emotional-motivational variables. Self-report instruments are prone to response biases (Kreitchmann et al., 2019), so emotional-motivational outcomes should be viewed as subjective indicators rather than objective evidence. Also, the study did not differentiate between intrinsic and extrinsic motivation. Future studies would benefit from more differentiated motivational instruments and qualitative data. Also aggregated metrics like mean scores are useful for identifying trends but may mask learner differences. Future research should include distributional analyses and subgroup comparisons to better understand variation across learners.

The tools differed substantially in their nature. The AI group used a dynamic, responsive chatbot with immediate feedback and contextual assistance, while the *Excel* group worked with a static spreadsheet without external help. This disparity likely influenced both performance and perceptions. *ExperiMentor* acted as a backend interpreter for Python-based queries, enabling interactive assistance that *Excel* could not match. Additionally, *ExperiMentor's* outputs were not systematically validated for scientific correctness during the intervention. While researcher oversight indicated plausible and coherent responses, it cannot be ruled out that misleading or incorrect information influenced some learners' interpretations. Further, observed motivational advantages may reflect novelty effects. Students may have rated the AI tool more favorably because of its newness. Recent studies suggest, that the sense of novelty regarding AI remains intact even after longer exposure (Long et al., 2024).

To overcome these limitations and build a more robust understanding of AI-mediated learning in physics education, future work could explore several directions:

Longitudinal studies could examine whether AI tool advantages are sustained across multiple exposures and whether skills developed with *ExperiMentor* transfer to real classroom teaching contexts. Multi-session interventions spanning several weeks or months would clarify whether motivational benefits persist and whether performance gains consolidate over time. Also, mixed-methods approaches combining quantitative performance measures with qualitative interviews and interaction log analyses could illuminate the source for motivational differences and reveal how learners engage with AI assistance. Diverse samples and contexts including in-service teachers and participants from multiple institutions and countries would strengthen generalizability claims. Comparative studies across different physics topics and educational levels would clarify boundary conditions for AI tool effectiveness.

Refined motivational measurement using instruments that differentiate intrinsic and extrinsic motivation, along with measures of self-efficacy, autonomy and competence, would provide more nuanced understanding of affective outcomes. Subgroup analyses examining how learners with different prior knowledge, attitudes toward technology or learning preferences respond to AI assistance would reveal for whom these tools work best. Further, design variations comparing *ExperiMentor* to other support strategies like human tutoring, enhanced static materials or hybrid approaches and testing different AI interaction styles would help isolate which features drive observed benefits. *ExperiMentor's* interactive feedback could be leveraged through gamification elements like achievement systems or progress visualization.

## 6 Conclusion

While both AI-supported (*ExperiMentor*) and traditional tools (*Excel*) facilitated measurable learning improvements, *ExperiMentor* led to significantly higher emotional and motivational engagement. However, these benefits did not translate into superior conceptual gains, suggesting that AI functions best as a complementary support rather than a standalone solution.

The observed engagement likely stems from the AI functioning as a More Knowledgeable Other, offering scaffolding that keeps learners within their Zone of Proximal Development. Furthermore, from a Cognitive Load Theory perspective, the tool's automation seems to have managed extraneous load (Sweller, 2012), while its interactivity satisfied Self-Determination Theory needs for autonomy and competence (Ryan and Deci, 2000). The AI-supported approach resulted in greater enjoyment, motivation, and perceived effectiveness. These results align with previous research indicating that AI tools promote engagement through interactive feedback and user-centered design through interactivity, adaptability and timely feedback (Popenici and Kerr, 2017; Liang et al., 2023; Farrokhnia et al., 2024). The observed motivation boost may partly reflect novelty effects or perceived usability advantages. This aligns with similar criticisms raised in educational research, such as by Buchner and Kerres (2023), who argue that media comparison studies often mask the conditions under which specific tools are effective. While the AI-based tool *ExperiMentor* showed motivational advantages, performance outcomes did not differ significantly from the *Excel* condition. This underscores the need to move beyond the question of whether AI "works better" than traditional tools, and toward understanding under what conditions, for which learners, and in what instructional designs AI tools provide added value.

These insights reinforce the idea that digital tools are not pedagogical strategies in and of themselves. Their effectiveness depends on how they are embedded within instructional settings, how much support they offer, and how well they align with educational goals. In this study, the AI and *Excel* groups not only differed in interface but also in the type and degree of feedback and guidance provided, which makes it difficult to isolate the effect of AI from the broader learning environment.

## Data availability statement

The original contributions presented in the study are included in the article/Supplementary material, further inquiries can be directed to the corresponding author.

## Ethics statement

Ethical approval was not required for this study, as the data collected were anonymized, and the survey did not involve sensitive or health-related topics. Participants were not exposed to any risk or burden, and the information gathered was and is not personally identifiable. Additionally, the survey was conducted solely for research purposes related to general study conditions, without any impact on participants' academic assessment, rights, or well-being. The studies were conducted in accordance with the local legislation and institutional requirements. The participants provided their written informed consent to participate in this study.

## Author contributions

JH: Data curation, Investigation, Conceptualization, Methodology, Writing – review & editing, Visualization, Writing – original draft, Formal analysis. JL: Investigation, Writing – review & editing, Data curation. SB-G: Supervision, Formal analysis, Writing – review & editing, Methodology, Investigation, Conceptualization. AB: Supervision, Software, Funding acquisition, Resources, Writing – review & editing.

## Funding

The author(s) declared that financial support was received for this work and/or its publication. The ComeMINT project was funded by the Federal Ministry of Education and Research as part of the BMBF funding line "Competence Centers for digital and digitally supported teaching in schools and continuing education in the STEM Field," funding code 01JA23M06G.

## Conflict of interest

The author(s) declared that this work was conducted in the absence of any commercial or financial relationships that could be construed as a potential conflict of interest.

The authors AB, SB-G declared that they were an editorial board member of Frontiers at the time of submission. This had no impact on the peer review process and the final decision.

## Generative AI statement

The author(s) declared that generative AI was used in the creation of this manuscript. This manuscript was linguistically developed and revised for better readability using OpenAI ChatGPT 4o, Anthropic Claude 3.7 Sonnet and the translation tool DeepL.

Any alternative text (alt text) provided alongside figures in this article has been generated by Frontiers with the support of artificial intelligence and reasonable efforts have been made to ensure accuracy, including review by the authors wherever possible. If you identify any issues, please contact us.

## Publisher's note

All claims expressed in this article are solely those of the authors and do not necessarily represent those of their affiliated organizations, or those of the publisher, the editors and the reviewers. Any product that may be evaluated in this article, or claim that may be made by its manufacturer, is not guaranteed or endorsed by the publisher.

## Supplementary material

The Supplementary Material for this article can be found online at: https://www.frontiersin.org/articles/10.3389/feduc.2026.1719670/full#supplementary-material